\documentclass[12pt]{JHEP3}
\pdfoutput=1
\usepackage{epsfig}
\usepackage{epstopdf}
\usepackage{graphicx}

\title{Gauss-Bonnet Black Holes and Heavy Fermion Metals}

\author{R. B. Mann\\ Department of Physics, University of Waterloo, 200 University
Avenue West, Waterloo, Ontario, Canada, N2L 3G1\\
E-Mail: \email{rbmann@sciborg.uwaterloo.ca}}
\author{R. Pourhasan\\ Department of Physics, University of Waterloo, 200 University
Avenue West, Waterloo, Ontario, Canada, N2L 3G1\\
E-Mail: \email{r2pourha@uwaterloo.ca}}

\abstract{We consider charged black holes in Einstein-Gauss-Bonnet Gravity with Lifshitz
boundary conditions.   We find that this class of models can reproduce the anomalous specific heat
of condensed matter systems exhibiting non-Fermi-liquid behaviour at
low temperatures.  We find that the temperature dependence of the Sommerfeld ratio is sensitive
to the choice of Gauss-Bonnet coupling parameter for a given value of the Lifshitz scaling parameter.
We propose that this class of models is dual to a class of models of non-Fermi-liquid systems proposed by Castro-Neto et.al. }
\begin{document}
\tableofcontents

\section{Introduction}

The AdS/CFT correspondence conjecture \cite{Maldacena1} has
motivated the development of dual gravity models to describe
strongly correlated systems in condensed matter physics (CMT)
\cite{Hartnoll, Rokhsar, Ardonne, Vishwanath}.   This ``AdS/CMT"
correspondence, while more speculative than its AdS/CFT
predecessor, has yielded some tantalizing insights into the
behaviour of these systems.    Systems of heavy fermions
\cite{hferm} -- in which electrons near the Fermi surface of
certain materials have an effective mass much larger than the free
electron mass -- are of particular interest since their behaviour
is not that of a conventional Fermi liquid.   The ratio of the
specific heat to temperature (the Sommerfeld ratio denoted by
$\gamma_0$), for low temperature, does not become constant but
instead rises as the temperature decreases \cite{Fliq}.
Indeed, there are some observations confirming that this
ratio has logarithmic temperature dependence \cite{NFL}. On the other hand, Castro Neto et al.
proposed a model which predicts that the physical properties of
the f-electron compounds at low temperatures are characterized by
weak power law behavior, $\gamma_0\equiv C/T\propto
T^{-1+\lambda}$ where $\lambda<1$ is some characteristic parameter
\cite{CNeto1, CNeto2}. This has been observed experimentally in a
number of non-Fermi liquid materials \cite{Maple1, Maple2}.
Though a full understanding of quantum criticality remains an
outstanding problem, it is believed that such non Fermi liquid
behaviour is due to the existence of a quantum critical point
\cite{qcrit}.

Near a critical point many condensed matter systems are described by  field theories with
anisotropic scaling symmetry of the form
\begin{equation}\label{scale}
t\rightarrow \lambda^{z}t,\qquad  \mathbf{x}\rightarrow
\lambda\mathbf{x},
\end{equation}
where $z(\geq 1)$ is a dynamical critical exponent represents the
degree of anisotropy between space and time; manifestly $z=1$
pertains to relativistic systems.  Applying the AdS/CMT conjecture entails consideration of spacetimes whose metrics
have the asymptotic form
\begin{equation}
ds^2=\ell^2\left(-r^{2z}dt^2+\frac{dr^2}{r^2}+r^2d\mathbf{x}^2\right)\label{asmet}
\end{equation}
where the coordinates ($t,r,x^{i}$) are dimensionless and the only
length scale in the geometry is $\ell $. Such spacetimes are conjectured to be the gravitational duals to field theories
with the scaling properties given in (\ref{scale}) \cite{Kachru}. It is straightforward to show that metric (\ref{asmet}) has the scaling properties in (\ref{scale}) provided $r\rightarrow\lambda^{-1}r$ as well.

Recently it was shown that these gravitational duals yield  non Fermi liquid behavior for the specific heat of  quantum critical system that is qualitatively similar to what is seen in experiments on heavy fermion systems \cite{Daniel1}. The effects
of finite temperature are induced by the presence of a black hole with the asymptotic form  (\ref{asmet}), and the addition of electromagnetic charge introduces another energy scale that can lead to interesting dynamics.  The specific heat
was found to scale as $T^{3/2}$ at high temperatures $T$, whereas at low $T$ the behaviour of the specific heat was sensitive to the scaling exponent $z$.  For $z=1$ conventional Fermi liquid behaviour was recovered (with
the Sommerfeld ratio becoming constant), whereas for
$z>1$ the Sommerfeld ratio increased with $T$ as is observed for several heavy fermion systems.

We consider in this paper the effects of adding in higher-order
curvature corrections in the form of a Gauss-Bonnet term on Fermi
liquid behaviour.   In the context of AdS/CFT correspondence,
inclusion of higher powers of the curvature is necessary in order
to understand CFTs with different values for their central
charges.  In the context of condensed matter holography, they
modify a number of important quantities, including viscosity, DC
conductivity, the superconducting phase transition, and more
\cite{GB1,GB2,GB3,PWang}. The simplest
and most natural such term to include in the action is the
Gauss-Bonnet term. Gauss-Bonnet gravity is a particular case of
Lovelock gravity, which in $(n+1)$ dimensions yields $p$-order
curvature terms (with $p\leq[n/2]$) in field equations with not
more than 2 derivatives of any metric function; terms for which $p
> [n/2]$ are total derivatives in the action and so do not
contribute to the field equations.   To obtain a (3+1)-dimensional
CMT, for which a (4+1) dimensional gravity theory is the
conjectured dual, it is therefore natural to include a
Gauss-Bonnet term on the gravity side since for $n=4$ we can only
have $p\leq 2$.

While we will ultimately be concerned with (3+1)-dimensional CMTs, we shall formulate our considerations in $(n+1)$ dimensions,
making use of the action
\begin{equation}
I=\int
d^{n+1}x\sqrt{-g}\left(R-2\Lambda+\hat{\alpha}\mathcal{L}_{\mathrm{GB}}
-\frac{1}{4}F_{\mu\nu}F^{\mu\nu}
-\frac{1}{4}H_{\mu\nu}H^{\mu\nu}-\frac{C}{2}B_{\mu}B^{\mu}\right),
\label{action}
\end{equation}
where $\Lambda $ is the cosmological constant and
$\mathcal{L}_{\mathrm{GB}}=R_{\mu\nu\gamma\delta}R^{\mu\nu\gamma\delta}-4R_{\mu\nu}R^{\mu\nu}+R^{2}$
is the second order Lovelock (Gauss-Bonnet) Lagrangian. The
coefficient $\hat{\alpha}$ has the dimension of (length)$^2$ and
can be positive or negative, $F_{\mu \nu }=\partial _{\lbrack \mu
}A_{\nu ]}$ with $A_{\mu }$ the Maxwell gauge field, and
$H_{\mu \nu }=\partial _{\lbrack \mu }B_{\nu ]}$ is the field
strength of the Proca field $B_{\mu }$ with mass $m^{2}=C$.

Neutral black hole solutions with $\alpha=0$ have been obtained
both analytically and numerically \cite{Daniel0,RM,Peet1}. The
$\alpha=0$ charged case in $(n+1)$ dimensions has been more
recently considered \cite{Daniel1, Daniel2, HRR}. In addition to
obtaining a holographic description of a strongly coupled quantum
critical point in 5 dimensions with asymmetric scaling that
exhibits the aforementioned anomalous specific heat behaviour at
low temperature for heavy fermion metals \cite{Daniel2}, a more
general study \cite{HRR} of the  $\hat{\alpha}=0$ version of
(\ref{action}) found a broad range of charged black hole solutions
and a general thermodynamic relationship between energy, entropy,
and chemical potential.  Solutions to  neutral Lovelock-Lifshitz
black holes were also recently obtained \cite{HR1} and their
thermodynamic behaviour for spatially flat cases was found to be
the same \cite{HR2} as the $p=1$ Einsteinian case \cite{Peet1}.
Perturbative corrections to the charged Lifshitz case due to a Gauss-Bonnet term ($\alpha \neq 0$) have been computed
\cite{pang}, but full black hole solutions and their implications
have not been studied.

We begin by introducing $(n+1)$-dimensional equations of motion in
sec. (\ref{FE}), where we find the constraints that must be
imposed on the constants in (\ref{action}) in order that the
asymptotic form (\ref{asmet}) hold. We then investigate solutions
at large $r$ in sec. (\ref{ls}), which provide us with useful
information about possible solutions due to different choices of
$\alpha$.  Near horizon expansions are computed in sec.
(\ref{nh}), and these are used to assist the numerical methods we
employ  to find  solutions. We analyze the thermodynamic behaviour
of these charged Gauss-Bonnet black holes numerically in sec.
(\ref{na}) where we specifically focus on $z=2$ charged black
holes in $(4+1)$ dimensions.  We find that the Gauss-Bonnet
coefficient $\alpha$ plays a role dual to that of the
characteristic parameter $\lambda$ in the Castro Neto et al. model
\cite{CNeto1, CNeto2}. In fact, if the coefficient $\alpha$ is
smaller than a certain minimal value $\alpha_m$, non-Fermi liquid
behaviour is enhanced, and the temperature dependence of the
Sommerfeld ratio $\gamma_0$ at low temperature can be
characterized as a weak power law.   However as $\alpha$ increases
above $\alpha_m$ the power law dependence of this ratio on the
temperature becomes vanishingly small, similar to the
$\lambda\rightarrow1$ limit in the Castro Neto et al. model,
which as they claim can be fit with the logarithmic behaviour
observed in some non fermi liquid compounds.  The Gauss-Bonnet
term restores Fermi liquid behaviour for $\alpha>>\alpha_m$.   We
propose  that Lifshitz-GB charged black holes are gravitational
duals to the Castro Neto et al. model. Further numerical
calculation reveals that the same qualitative behaviour happens
for arbitrary $z >1$. We finish our paper by giving some closing
remarks pertinent to our results.

\section{Field Equations in ($n+1$)-dimensions}
\label{FE}

Using the variational principal the field equations that follow
from the action (\ref{action}) are:
\begin{eqnarray}
&&G_{\mu\nu}+\hat{\alpha}G_{\mu\nu}^{\mathrm{GB}}+\Lambda
g_{\mu\nu}=T_{\mu\nu},\label{EqG}\\
&&\nabla^{\mu}H_{\mu\nu}=CB_{\mu},\label{EqH}\\
&&\partial_{[\mu}B_{\nu]}=H_{\mu\nu},\label{EqB}\\
&&\nabla^{\mu}F_{\mu\nu}=0,\label{Eqk}
\end{eqnarray}
where
\begin{equation}
T_{\mu\nu}=-\frac{1}{2}\left(\frac{1}{4}F_{\rho\sigma}F^{\rho\sigma}g_{\mu\nu}-F_{\phantom{\rho}{\mu}}^{\rho}
F_{\rho\nu}+\frac{1}{4}H_{\rho\sigma}H^{\rho\sigma}g_{\mu\nu}-H_{\phantom{\rho}{\mu}}^{\rho}
H_{\rho\nu}+C\left[\frac{1}{2}B_{\rho}B^{\rho}g_{\mu\nu}-B_{\mu}B_{\nu}\right]\right)
\end{equation}
is the energy-momentum tensor of gauge fields, $G_{\mu\nu}$ is the
Einstein tensor, and $G_{\mu\nu}^{\mathrm{GB}}$ is given as:
\begin{eqnarray}
G_{\mu\nu}^{\mathrm{GB}}&=&2(-R_{\mu\sigma\kappa\tau}R_{\phantom{\kappa\tau\sigma}{\nu}}^{\kappa\tau\sigma}%
-2R_{\mu\rho\nu\sigma}R^{\rho\sigma}-2R_{\mu\sigma}R_{\phantom{\sigma}{\nu}}^{\sigma}+RR_{\mu\nu})-\frac{1}{2}\mathcal{L}_{\mathrm{GB}}g_{\mu\nu},
\end{eqnarray}
The $(n+1)$-dimensional metric that preserves the basic symmetries
under consideration can be written as
\begin{equation}
ds^{2}=\ell^2\left(-r^{2z}f^{2}(r)dt^{2}+\frac{g^{2}(r)dr^{2}}{r^{2}}+r^{2}d\mathbf{x}^{2}\right) \label{metric}
\end{equation}
which is that of a black brane or (if appropriate identifications are carried out) a toroidal black hole.

The gauge fields are assumed to be
\begin{equation}
A_{t}=\ell r^{z}\kappa(r),\qquad B_{t}=q\ell r^{z}f(r)j(r),\qquad
H_{tr}=q\ell zr^{z-1}g(r)h(r)f(r), \label{gaugeans}
\end{equation}
with all other components either vanishing or being given by
antisymmetrization. In order to get Lifshitz geometry
(\ref{asmet}) as the asymptotic form of the metric (\ref{metric}),
one should demand $f(r)=g(r)=h(r)=j(r)=1$ and $k(r)=0$ as $r$ goes
to infinity which impose the following constraints over the
constants:
\begin{eqnarray}
&&C=\frac{(n-1)z}{\ell^{2}},\qquad
q^{2}=\frac{2(z-1)L^{2}}{z\ell^{2}},\nonumber\\
&&\Lambda_{\mathrm{eff}}=-\frac{\left[(z-1)^{2}+n(z-2)+n^{2}\right]L^2%
+n(n-1)\alpha}{2\ell^{4}},\label{cons}
\end{eqnarray}
with the following redefinition
\begin{equation}
L^{2}=\ell^2-2\alpha,\qquad \alpha=(n-2)(n-3)\hat{\alpha}
\end{equation}
where according to eq. (\ref{cons}) $L^2$ should be positive which
implies $\alpha<\ell^2/2$ and consequently the cosmological
constant $\Lambda_{\mathrm{eff}}$ is always negative.
Note that upon setting $z=1$ in (\ref{cons}) we recover
an effective cosmological constant dependent on the coupling $\alpha$
rather than the usual AdS cosmological
constant, i.e. $\Lambda=-n(n-1)/2\ell^2$.

Applying the ansatz (\ref{metric}) to the equation (\ref{Eqk})
yields the solution
\begin{equation}
(r^z\kappa)^{\prime}=\frac{Q}{r^{n-z}}fg \label{dk}
\end{equation}
where $Q$ is an  integration constant related to the Maxwell
charge and we have chosen boundary conditions such that the
Maxwell vector potential vanishes at the horizon.

Substituting (\ref{cons}) and (\ref{dk}) into eqs.
(\ref{EqG}-\ref{EqB}), the field equations reduce to the following system of
first order differential equations:
\begin{eqnarray}
r\frac{df}{dr}&=&\frac{f}{4(n-1)(\ell^2g^2-2\alpha)}\{2\left[(n-1)(z-1)j^{2}-z(z-1)h^{2}+
(z-1)^{2}+n(z-2)+n^{2}\right]g^{4}L^2\nonumber\\
&+&2\alpha(n-1)(ng^{4}+n+4z-4)-2(n-1)(n+2z-2)\ell^{2}g^{2}-Q^{2}r^{2-2n}g^{4}\},\label{Eqf}\\
r\frac{dg}{dr}&=&\frac{g}{4(n-1)(\ell^2g^2-2\alpha)}\{2\left[(n-1)(z-1)j^{2}+z(z-1)h^{2}-
(z-1)^{2}-n(z-2)-n^{2}\right]g^{4}L^2\nonumber\\
&-&2\alpha n(n-1)(1+g^{4})+2n(n-1)\ell^2g^2+Q^{2}r^{2-2n}g^{4}\},\label{Eqg}\\
r\frac{dj}{dr}&=&-\frac{j}{4(n-1)(\ell^2g^2-2\alpha)}\{2\left[(n-1)(z-1)j^{2}-z(z-1)h^{2}+
(z-1)^{2}+n(z-2)+n^{2}\right]g^{4}L^2\nonumber\\
&+&2\alpha
(n-1)(ng^{4}+n-4)-2(n-1)(n-2)\ell^{2}g^{2}-Q^{2}r^{2-2n}g^{4}\}+zgh,\label{Eqj}\\
r\frac{dh}{dr}&=&(n-1)(jg-h).\label{Eqh}
\end{eqnarray}

\section{Series Solutions}

\subsection{Solutions at large $r$}
\label{ls}

Since we wish to  understand the effects of the Gauss-Bonnet term
on the previously considered \cite{HRR}   ($n+1$)-dimensional charged Einstein solutions with
arbitrary $z$, we begin by  linearizing the system in ($n+1$)-dimensions.
Requiring the general metric (\ref{metric}) to asymptotically
approach the Lifshitz one, we write
\begin{eqnarray}
&&f(r)=1+w f_1(r),  \nonumber \\
&&g(r)=1+w g_1(r),  \nonumber \\
&&j(r)=1+w j_1(r),  \nonumber \\
&&h(r)=1+w h_1(r)   \label{infexh}
\end{eqnarray}
where we note that it is also necessary \cite{Daniel0,RM} for $j(r)$ and $h(r)$ to approach unity in order to obtain (\ref{asmet}).

In previous section, we already introduced equations of motion
which are shown in (\ref{Eqf}-\ref{Eqh}). Since $f(r)$ does not
contribute to the equations for $g(r)$, $h(r)$ and $j(r)$ let's
first study a set involving $\{g,h,j\}$; If one inserts
perturbative expansion (\ref{infexh}) into equations
(\ref{Eqg}-\ref{Eqh}) one obtains the equations for small
perturbations as
\begin{eqnarray}
r\frac{d}{dr}\pmatrix{ g_1\cr h_1\cr
j_1}&=&\pmatrix{n&&-z(z-1)/(n-1)&&1-z\cr1-n&&n-1&&1-n\cr
M_{31}(n,z,\alpha)&&-z(n+z-2)/(n-1)&&2z-1}\pmatrix{
g_1\cr h_1\cr j_1} \nonumber \\
&&\qquad +\frac{Q^2}{4(n-1)r^{2n-2}}
\pmatrix{1\cr0\cr1}\label{Eqp}
\end{eqnarray}
which is identical to the Einsteinian case except for the term
\begin{equation}
M_{31}\equiv\frac{(n+z-2)\ell^{2}-2(n+3z-4)\alpha}{L^{2}}
\end{equation}
where we have rescaled $r\rightarrow r/\ell$.   We have also included the Maxwell gauge field as a first order
perturbation, which means that we substitute
$Q^2/r^{2(n-1)}=wQ^2/r^{2(n-1)}$, since its falloff may be slower
than other terms in the metric functions.

Solving the set (\ref{Eqp}) one obtains:
\begin{eqnarray}
&&g_1(r)=-\frac{C_{1}G_{1}}{r^{z+n-1}}-\frac{C_{2}G_{2}}{r^{(z+n-1+\sqrt{%
\gamma})/2}}-\frac{C_{3}G_{3}}{r^{(z+n-1-\sqrt{\gamma})/2}}-\frac{(n-2z)\ell^2Q^2}{4\Delta(n-z-1)r^{2n-2}},  \label{glr} \\
&&h_1(r)=-\frac{C_{1}}{r^{z+n-1}}-\frac{C_{2}}{r^{(z+n-1+\sqrt{\gamma})/2}}%
-\frac{C_{3}}{r^{(z+n-1-\sqrt{\gamma})/2}} \nonumber\\
&&\qquad\qquad\qquad\qquad\qquad+\frac{[(2n-z-2)\ell^2-2(2n-3)\alpha]\ell^2Q^2}{4L^2\Delta(n-z-1)r^{2n-2}},  \label{hlr} \\
&&j_1(r)=-\frac{C_{1}J_{1}}{r^{z+n-1}}-\frac{C_{2}J_{2}}{r^{(z+n-1+\sqrt{%
\gamma})/2}}-\frac{C_{3}J_{3}}{r^{(z+n-1-\sqrt{\gamma})/2}} \nonumber\\
&&\qquad\qquad\qquad\qquad\qquad -\frac{[(n+z-2)\ell^2-2(n+2z-3)\alpha]\ell^2Q^2}{4L^2\Delta(n-z-1)r^{2n-2}},
\label{jlr}
\end{eqnarray}
where
\begin{eqnarray}
&&\gamma=\frac{((n-3z)^2+6n-2z-7)\ell^2-2(n^2-6nz+6n+14z+z^2-15)\alpha}{L^2}, \\
&&\Delta=(n+z-2)(n-z-1)\ell^2-2(n^2-3n-z+3)\alpha\\
&&G_{1}=\frac{z(z-1)L^2}{(n-1)^2\ell^2-2(n-1)(n+z-2)\alpha},\quad G_{2}=\frac{z-1}{n-1},\quad G_{3}=\frac{%
z-1}{n-1}, \\
&&J_{1}=-\frac{z(n+z-2)\ell^2-2(n+2z-3)\alpha}{(n-1)^2\ell^2-2(n-1)(n-z+2)\alpha},\\
&&J_{2}=\frac{n-3z+1-\sqrt{\gamma}%
}{2(n-1)},\quad
J_{3}=\frac{n-3z+1+\sqrt{\gamma}}{2(n-1)}.
\end{eqnarray}
Substituting above expressions into the equation for  small
perturbations of $f(r)$
\begin{eqnarray}
r\frac{d}{dr}f_1(r)&=&\frac{(n-2+2z)\ell^2-2(n+4z-4)\alpha}{L^2}g_1(r)-\frac{z(z-1)}{n-1}
h_1(r)\nonumber\\&+&(z-1)j_1(r)-\frac{\ell^2Q^2}{4L^2(n-1)r^{2n-2}}
\end{eqnarray}
we obtain
\begin{eqnarray}
f_1(r)&=&-\frac{C_{1}F_{1}}{r^{z+n-1}}-\frac{C_{2}F_{2}}{r^{(z+n-1+\sqrt{%
\gamma})/2}}-\frac{C_{3}F_{3}}{r^{(z+n-1-\sqrt{\gamma})/2}} \nonumber\\
&&\qquad\qquad\qquad\qquad\qquad +\frac{[(n+z-2)\ell^2-2(n+2z-3)\alpha]\ell^2Q^2}{4L^2\Delta(n-1)r^{2n-2}},
\label{flr}
\end{eqnarray}
with
\begin{eqnarray}
&&F_1=\frac{(n+2z-2)\ell^2-2(n+4z-4)\alpha}{(n+z-1)L^2}G_1+\frac{z-1}{n+z-1}J_1\nonumber\\
&&\qquad\qquad\qquad\qquad\qquad -\frac{z(z-1)}{(n-1)(n+z-1)},\\
&&F_2=\frac{(n+2z-2)\ell^2-2(n+4z-4)\alpha}{(n+z-1+\sqrt{\gamma})L^2}G_2+\frac{2(z-1)}{n+z-1+\sqrt{\gamma}}J_2
\nonumber\\
&&\qquad\qquad\qquad\qquad\qquad -\frac{2z(z-1)}{(n-1)(n+z-1+\sqrt{\gamma})},\\
&&F_3=\frac{(n+2z-2)\ell^2-2(n+4z-4)\alpha}{(n+z-1-\sqrt{\gamma})L^2}G_3+\frac{2(z-1)}{n+z-1-\sqrt{\gamma}}J_3 \nonumber\\
&&\qquad\qquad\qquad\qquad\qquad  -\frac{2z(z-1)}{(n-1)(n+z-1-\sqrt{\gamma})}.
\end{eqnarray}

There are three eigenmodes, of which two are decaying. The third
one might be decaying, growing or independent of $r$ depending on
the value of $\alpha$.  There is a zero mode if
\begin{equation}
\alpha^{(0)}=\frac{(n-z-1)\ell^2}{2(n-2)}.
\end{equation}
Accordingly, for $\alpha\geq\alpha^{(0)}$ in order to approach the
Lifshitz fixed point asymptotically,  fine-tuning of initial
values is required when numerically solving the non-linear field
equations. However  if $\alpha<\alpha^{(0)}$, all three eigenmodes
are decaying, yielding a family of solutions with the same event
horizon that asymptotically approach the Lifshitz background with
different fall-off rates.

\subsection{Near horizon expansion}
\label{nh}

To investigate the near horizon behavior of the solutions
we  consider the following expansions
\begin{eqnarray}
&&f(r)=f_{0}\sqrt{r-r_{0}}(1+f_{1}(r-r_{0})+f_{2}(r-r_{0})^{2}+\cdots),\nonumber\\
&&g(r)=\frac{g_{0}}{\sqrt{r-r_{0}}}(1+g_{1}(r-r_{0})+g_{2}(r-r_{0})^{2}+\cdots),\nonumber\\
&&j(r)=j_{0}\sqrt{r-r_{0}}(1+j_{1}(r-r_{0})+j_{2}(r-r_{0})^{2}+\cdots),\nonumber\\
&&h(r)=h_{0}(1+h_{1}(r-r_{0})+h_{2}(r-r_{0})^{2}+\cdots),\label{nhexp}
\end{eqnarray}
where $r_{0}$ is assumed to be the horizon radius. Inserting this
into Eqs. (\ref{Eqf}-\ref{Eqh}) and demanding the coefficients for
each power of $(r-r_{0})$  vanish  yields  relations
between the various constant coefficients in terms of $r_{0}$ and
$h_{0}$. For example $g_{0}$ is given by:
\begin{equation}
g_{0}=\ell r_{0}^{1/2}\sqrt{\frac{2(n-1)}
{\ell^{2}r_{0}^{2-2n}(Q_{\mathrm{ext}}^{2}-Q^{2})-2z(z-1)L^{2}h_{0}^{2}}}\label{g0}
\end{equation}
where
\begin{eqnarray}
Q_{\mathrm{ext}}^{2}\equiv\frac{2}{\ell^{2}r_{0}^{2-2n}}\left%
[\left(n^{2}+(z-2)n+(z-1)^{2}\right)\ell^{2}-\left(n^{2}+(2z-3)n+2(z-1)^{2}\right)\alpha\right]
\label{Qext}
\end{eqnarray}
is the upper bound for the Maxwell charge for a given $\alpha$.
Since $\alpha<\ell^2/2$ the right hand side of eq. (\ref{Qext}) is
always positive.   Clearly $Q_{\mathrm{ext}}$ decreases as
$\alpha$ increases. Furthermore, from (\ref{g0}) we see that real
series solutions exist provided
\begin{equation}
h_{0}<\sqrt{\frac{\ell^{2}r_{0}^{2-2n}(Q_{\mathrm{ext}}^{2}-Q^{2})}{2z(z-1)L^{2}}}
\end{equation}
with $Q\leq Q_{\mathrm{ext}}$.
All other constants in the series solutions (\ref{nhexp}) can be
obtained in terms of $r_{0}$ and $h_{0}$ as well but we don't
write them here because they are too lengthy.

\section{Numerical analysis and Thermodynamic Behaviour}
\label{na}

To find numerical solutions for the system of ODE's
(\ref{Eqf}-\ref{Eqh}) we apply the shooting method, adjusting
initial values for the fields $f(r)$, $g(r)$, $j(r)$ and
$h(r)$ and requiring each to approach their asymptotic values of unity.
 To find these initial values we use the  near horizon
expansions (\ref{nhexp}) by substituting $r_{0}+\varepsilon$ for
$r$ where $\varepsilon\ll1$ and choose some values for $r_{0}$ and
$h_{0}$ such that  the fields approach unity at large $r$ to
within a certain tolerance of $10^{-8}$.  While we can obtain explicit numerical solutions for the
metric functions, we are interested here in their thermodynamic
behaviour, particularly insofar as how
 the Gauss-Bonnet term modifies the Sommerfeld relation.
Throughout we restrict our numerical analysis to 5 dimensions.

\subsection{Thermal behaviour}

The Hawking temperature of a general black brane (\ref{metric}) is
\begin{equation}
T=\frac{r_0^{z}}{4\pi}\tau_z(r_0,Q,\alpha), \label{temp}
\end{equation}
with
\begin{equation}
\tau_z=\frac{r_0f_0}{g_0}, \qquad z\neq1
\end{equation}
where $f_0$ and $g_0$ are ($r_0,\,Q,\,\alpha$)-dependant
coefficients from the near horizon expansions (\ref{nhexp}). Noting from eq. (\ref{Eqf}) that there is
rescaling symmetry for $f(r)$, we write $f_0\rightarrow \widetilde{f}_0/\sqrt{r_0}$.
Defining $\widetilde{Q}\equiv Q/r_0^{n-1}$ we find numerically
that $\tau_z$ is just a function of $(\widetilde{Q},\alpha)$ and does not depend on $r_0$
explicitly.

To examine the thermal behaviour of these charged Gauss-Bonnet
black branes we  follow two different approaches. First we
investigate how temperature depends on the Gauss-Bonnet
coefficient $\alpha$ for fixed electric charge. Then we
investigate the variation of temperature with electric charge for
fixed Gauss-Bonnet coefficient.

\begin{figure}[tbp]
\centering
{\includegraphics[width=.4\textwidth]{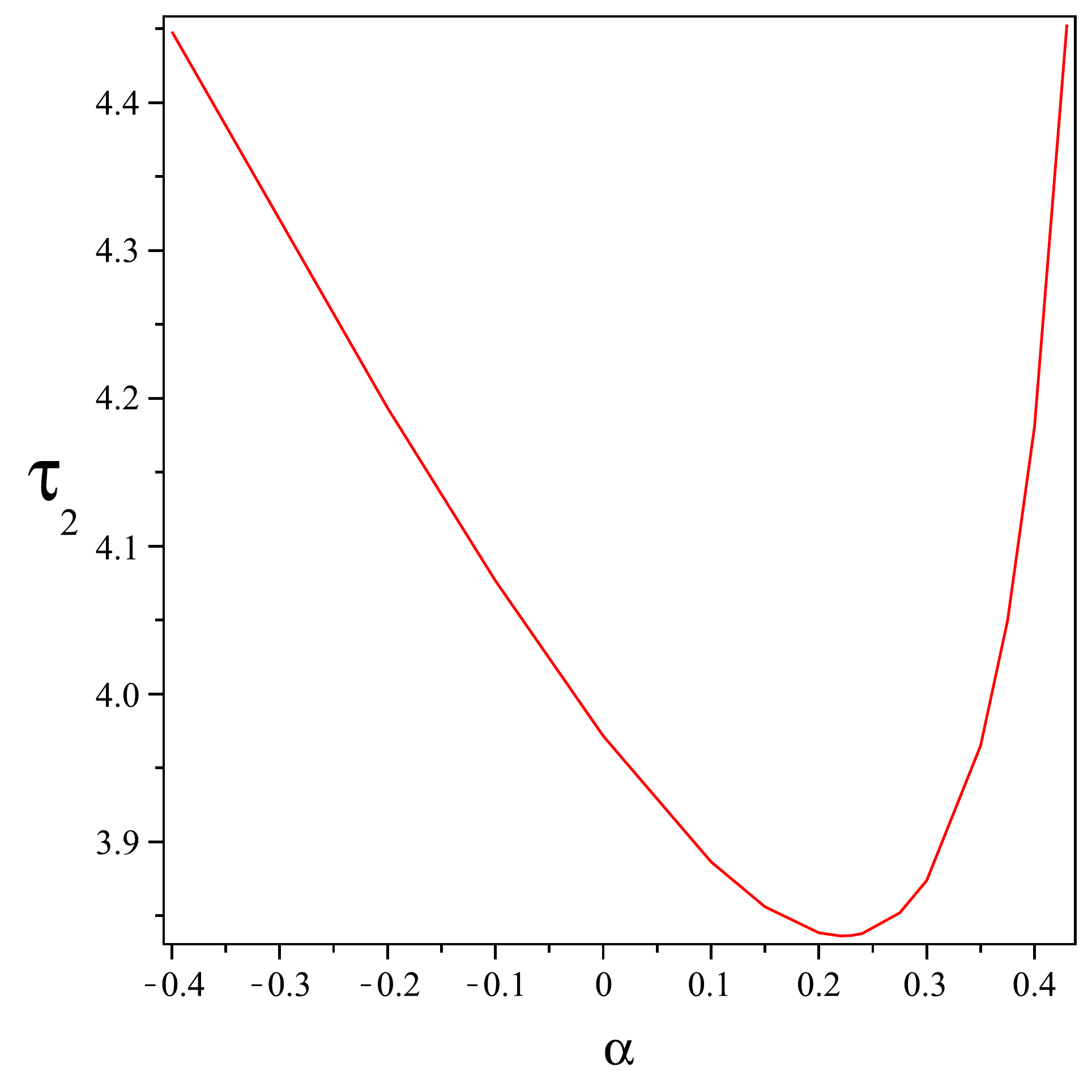}\qquad}
{\includegraphics[width=.4\textwidth]{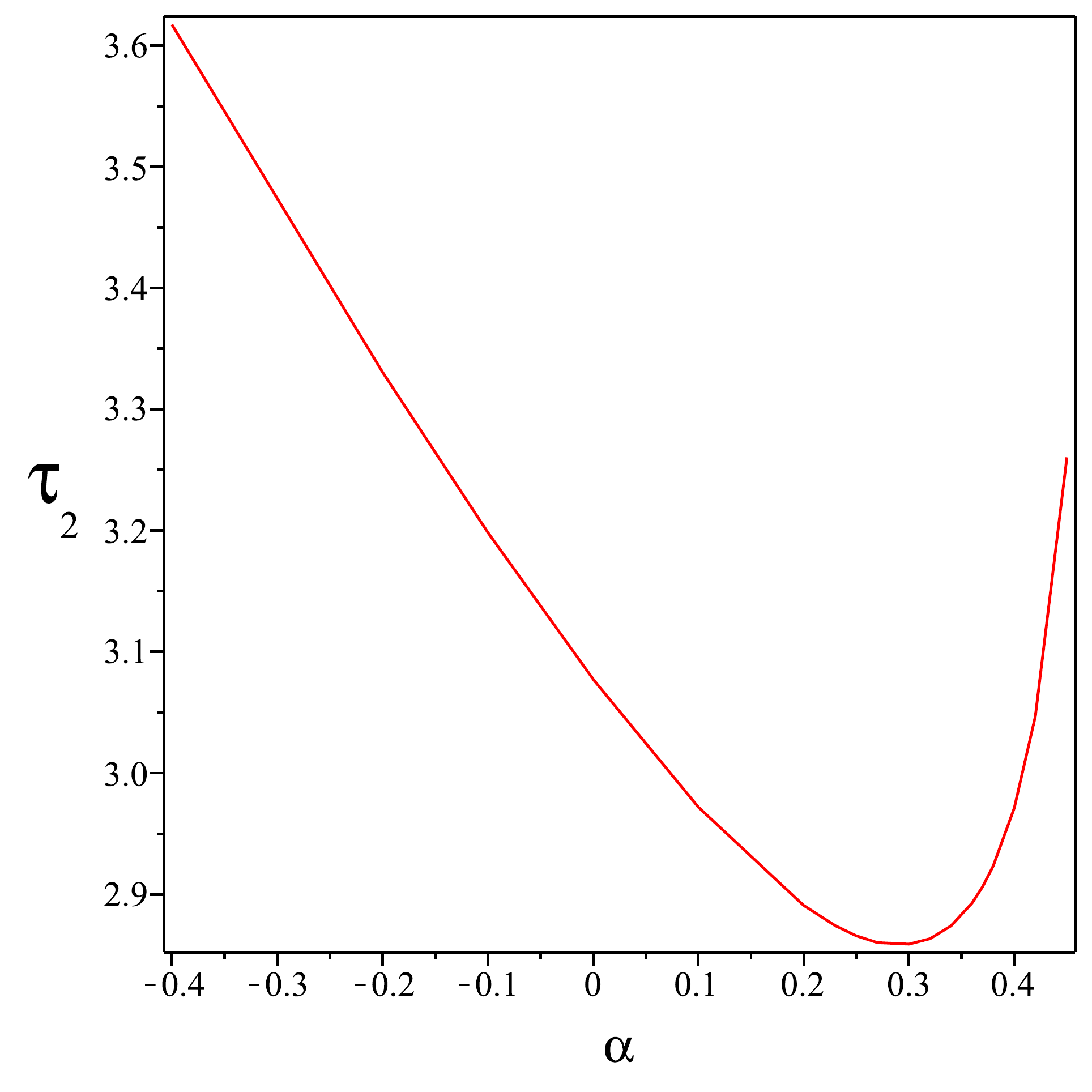}}
\caption{$\tau_2$ vs $\alpha$ in 5-dimensions with $z=2$. Left:
neutral black holes with $\alpha_m\approx0.22$. Right: weakly
charged black holes ($\widetilde{Q}\approx0.676\,
\widetilde{Q}_c$) with $\alpha_m\approx0.3$. We have set $\ell
=1$.} \label{tvsaz2}
\end{figure}

Consider first fixed electric charge $\widetilde{Q}$. For $z=1$
the temperature function $\tau_1$ depends linearly on $\alpha$ as
\begin{equation}
\tau_1=n(1-\frac{\alpha}{\ell^2})-\frac{\widetilde{Q}^2}{2(n-1)},\label{tempz1}
\end{equation}
while numerical analysis indicates that the situation is different
for $z>1$. The reality condition from eq. (\ref{g0}) imposes the
constraint $\alpha\leq\alpha_{\mathrm{ext}}$, where
\begin{equation}
\alpha_{\mathrm{ext}}=\frac{2n^{2}+2(z-2)n+2(z-1)^{2}-\widetilde{Q}^2}{2n^{2}+2(2z-3)n+4(z-1)^{2}}\ell^{2}
\quad . \label{aext}
\end{equation}
Recalling that   $\alpha<\ell^2/2$, we define
$\widetilde{Q}_{c} = \widetilde{Q}(\alpha_{\mathrm{ext}}=\ell^2/2)$ from
(\ref{aext}), yielding
\begin{equation}
\widetilde{Q}_{c}^{2} \equiv n(n-1). \label{Qc}
\end{equation}
For charged black holes with $\widetilde{Q}<\widetilde{Q}_c$ we
have $\alpha_{\mathrm{ext}}>\ell^2/2$ implying the bound $\alpha \leq \ell^2/2$;
otherwise, if $\widetilde{Q}_c < \widetilde{Q} <
\widetilde{Q}_{\mathrm{ext}}(=Q_{\mathrm{ext}}/r_0^{n-1})$ then
$\alpha < \alpha_{\mathrm{ext}} < \ell^2/2$.

Numerically solving the field equations, we find that weakly charged ($\widetilde{Q}<\widetilde{Q}_c$) or neutral
black holes  with $z>1$ get colder as $\alpha$
decreases from its maximal value of $\ell^2/2$ down to a specific
value, denoted $\alpha_m$, which can be determined numerically. As
$\alpha$ further decreases below $\alpha_m$ the temperature starts
increasing. In other words the temperature is minimized at $\alpha
= \alpha_m < \ell^2/2$.

\begin{figure}[tbp]
\centering
{\includegraphics[width=.4\textwidth]{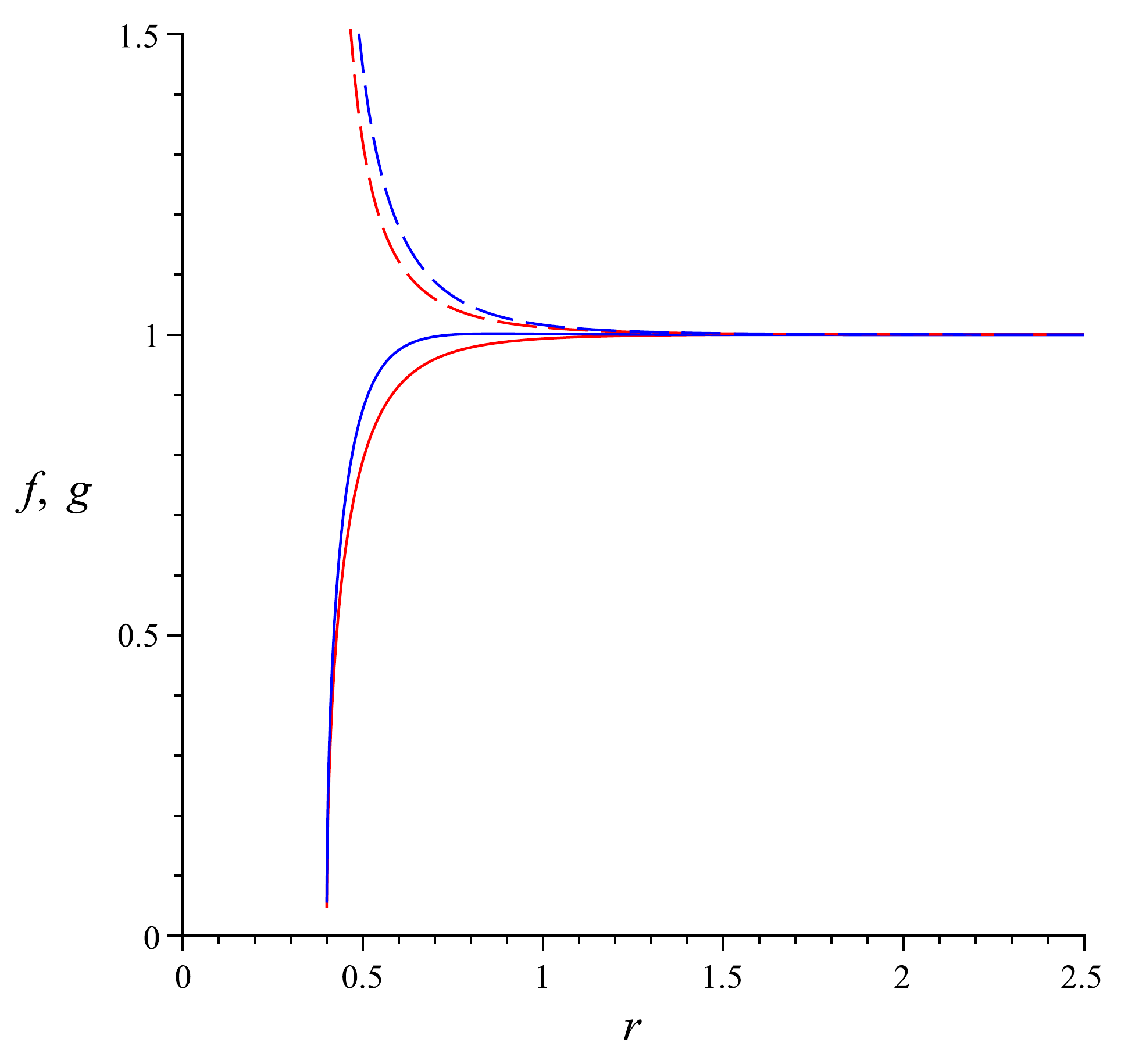}\qquad} {%
\includegraphics[width=.4\textwidth]{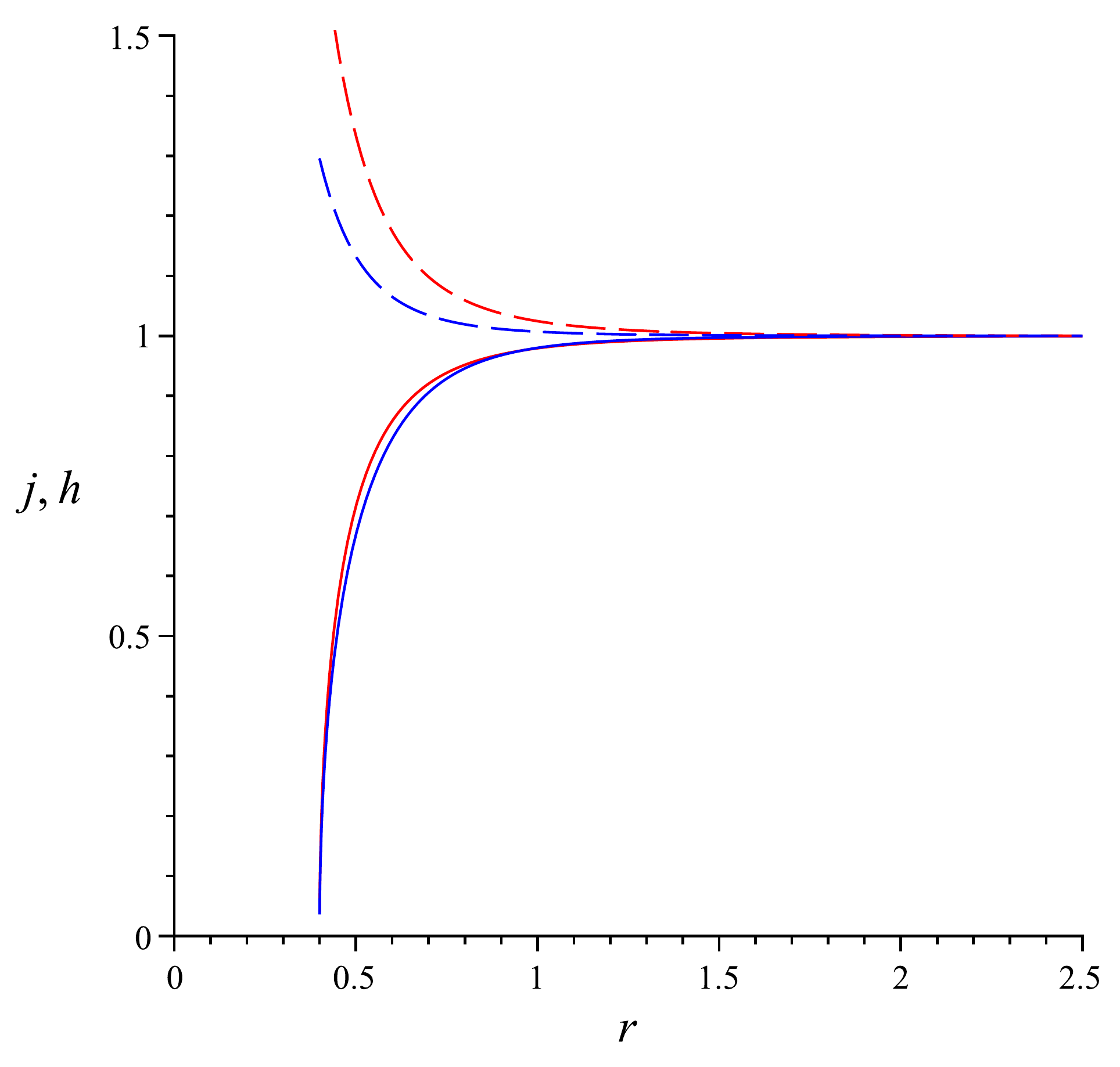}}
\caption{Functions describing 5-dimensional neutral
Einstein ($\widetilde{Q}=0,\,\alpha=0$) in red lines and neutral
Gauss-Bonnet($\widetilde{Q}=0,\,\alpha=0.36$) in blue lines both
with $z=2$. Left: metric functions $f(r)$(solid) and
$g(r)$(dashed). Right: gauge functions $j(r)$(solid) and
$h(r)$(dashed).   We have set $\ell =1$.} \label{frEGQ0}
\end{figure}

We depict this behaviour in  fig. \ref{tvsaz2} for small and
large black holes in 5 dimensions with $z=2$.   It is evident from
fig. \ref{tvsaz2} that it is possible to find isothermal pairs
of black holes. For example, a $z=2$ neutral Einstein black
hole ($\alpha=0$) with $r_0=0.4$ is isothermal to a neutral GB
black hole with $\alpha=0.36$ of the same $z$ and size. We
 illustrate in fig.  \ref{frEGQ0}  that these are
indeed two different solutions.

\begin{figure}[tbp]
\centering{\includegraphics[width=.4\textwidth]{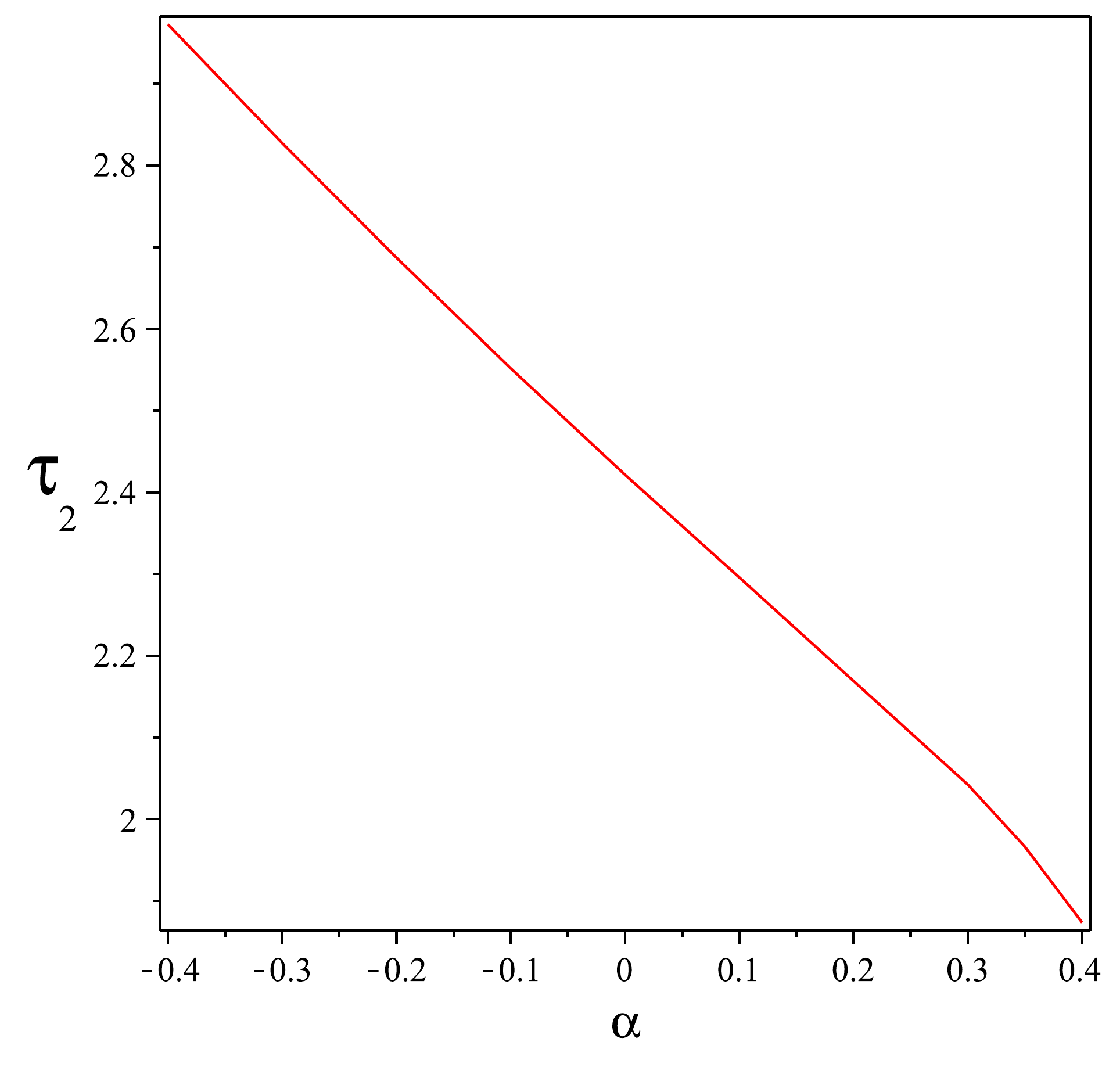}\qquad}
\centering{\includegraphics[width=.4\textwidth]{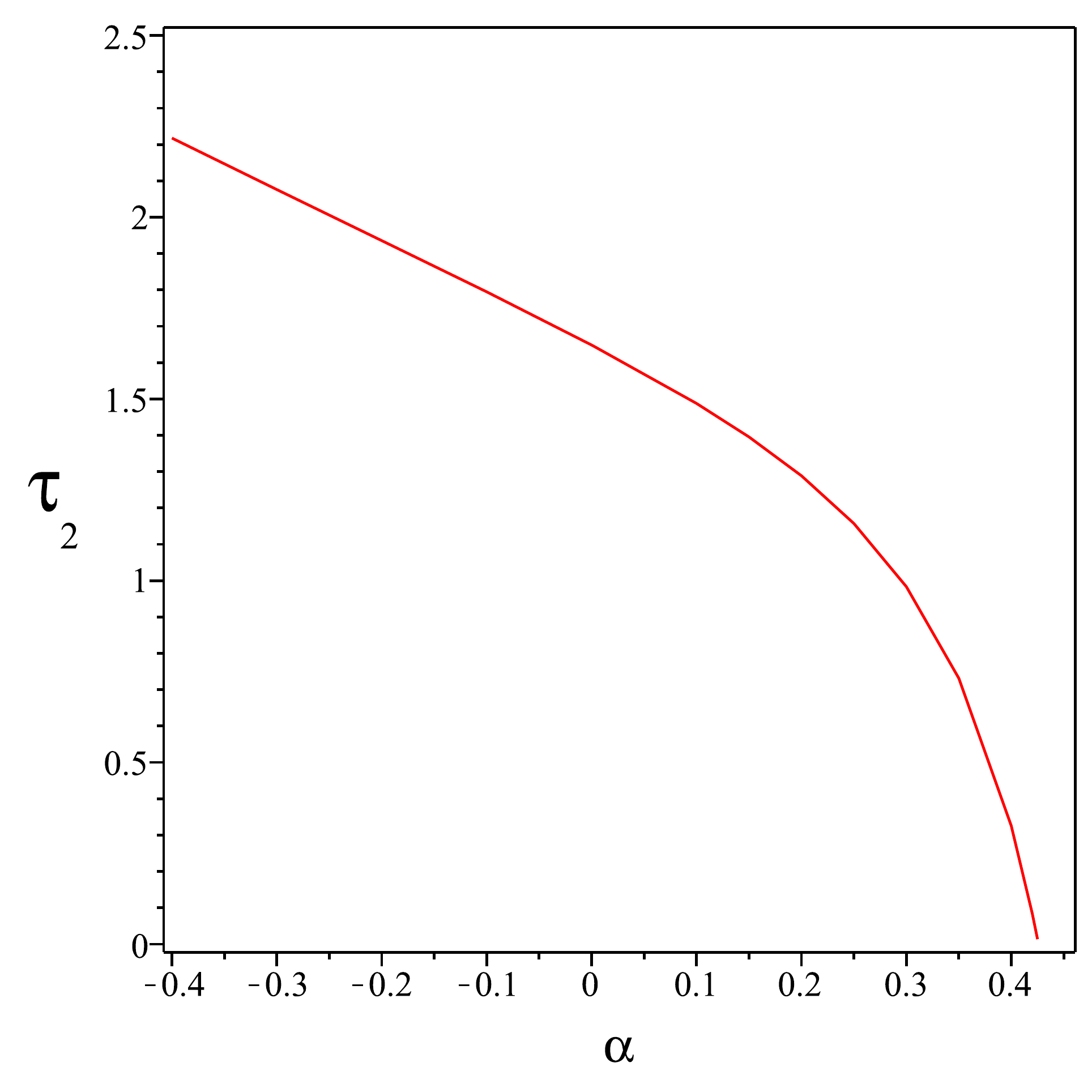}}
\caption{$\tau_2$ vs $\alpha$ in 5-dimensions with $z=2$.
Left: intermediate charged black holes ($\widetilde{Q}\approx
0.902\,\widetilde{Q}_c$) with $\alpha_{\mathrm{ext}}\approx0.55$.
Right: strongly charged black holes
($\widetilde{Q}\approx1.128\,\widetilde{Q}_c)$ with
$\alpha_{\mathrm{ext}}\approx0.425$.  We have set $\ell =1$.}
\label{tvsaz2Q}
\end{figure}

For charged black holes with intermediate values of the electric
charge  -- $\widetilde{Q} \lesssim \widetilde{Q}_c$ -- we find
that the temperature has a minimum nonzero value at $\alpha
\lesssim \ell^2/2$ and then increases monotonically as $\alpha$
decreases from $\ell^2/2$. There are no isothermal pairs. We
illustrate this behaviour for typical cases in the left-hand side
of fig. \ref{tvsaz2Q}.

The temperature $T$ of  strongly charged GB black holes, where
$\widetilde{Q}>\widetilde{Q}_c$ (and $z>1$) exhibits markedly
different behaviour. In this case  $\alpha <
\alpha_{\mathrm{ext}}<\ell^2/2$ and we find, as shown in the
right-hand side of fig. \ref{tvsaz2Q}, that $T$ starts from zero
at $\alpha= \alpha_{\mathrm{ext}}$ and then monotonically
increases with decreasing $\alpha$.  Again, no isothermal pairs
occur.

Turning next to the question of how temperature changes
with charge for fixed  $\alpha$, we shall consider separately the
 high temperature and low temperature behaviour.

For a given $\alpha$, high temperature behaviour corresponds to
$\widetilde{Q}\rightarrow0$. In this limit  the temperature
function $\tau_z$ is just a function of $\alpha$ and independent
of $r_0$. Hence $ T \propto r_0^z$ for charged GB black holes at
high temperature with arbitrary $z$.

\begin{figure}[tbp]
\centering {\includegraphics[width=.9\textwidth]{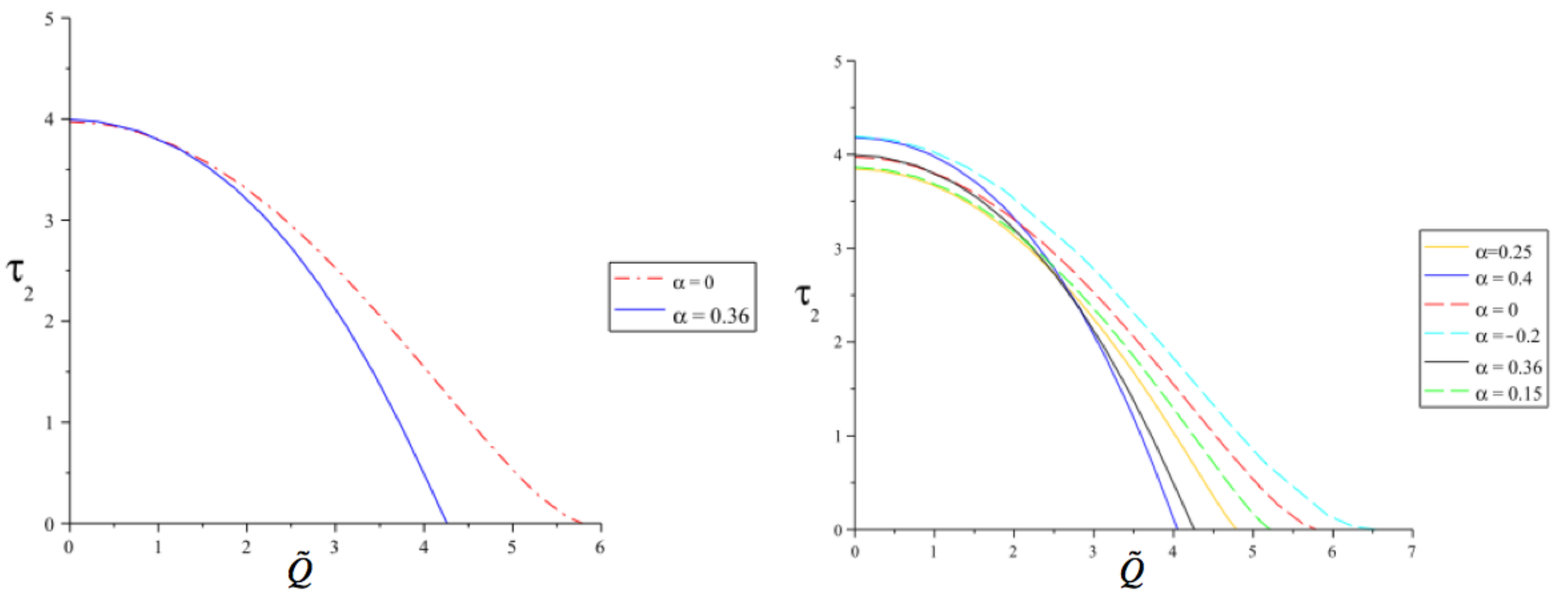}}
\caption{$\tau_2$ vs $\widetilde{Q}$ for $z=2$ black holes in 5
dimensions with $\alpha_m\approx0.22$. In the right hand plot
solids show $\alpha>\alpha_m$ where dashed show
$\alpha<\alpha_m$. } \label{tvsQz2}
\end{figure}

We have already noted that in the high temperature limit there
exist pairs of isothermal black holes with different GB
coefficients, one with $\alpha>\alpha_m$ and the other with
$\alpha<\alpha_m$.  (The minimum GB coefficient
$\alpha_m\approx0.22,\, 0.02$ and $-0.1$ for $z=2,\,3$ and $4$
respectively). We consider two such isothermal cases in  fig.
\ref{tvsQz2}. We see that for small $\widetilde{Q}$ (or high
temperature) the plots for the isothermal pairs $\alpha=(0,0.36)$
and $\alpha=(-0.2,0.4)$ (where $\alpha_m \simeq 0.22$) are
essentially the same.

However the distinction between pairs grows with increasing
$\widetilde{Q}$.  As the low temperature limit
($\widetilde{Q}\rightarrow\widetilde{Q}_{\mathrm{ext}}$)  is
approached, the behaviour of $\tau_z$  becomes significantly
different between the elements of the isothermal pair. For
$\alpha<\alpha_m$ the low temperature limit ($T\rightarrow0$) is
non-linear as a function of increasing $\widetilde{Q}$, whereas
for its isothermal counterpart with $\alpha>\alpha_m$, it tends
toward linearity. We will revisit this distinction in the next
section where we will discuss the behaviour of the specific heat
at low temperature.

\begin{figure}[tbp]
\centering
{\includegraphics[width=.34\textwidth]{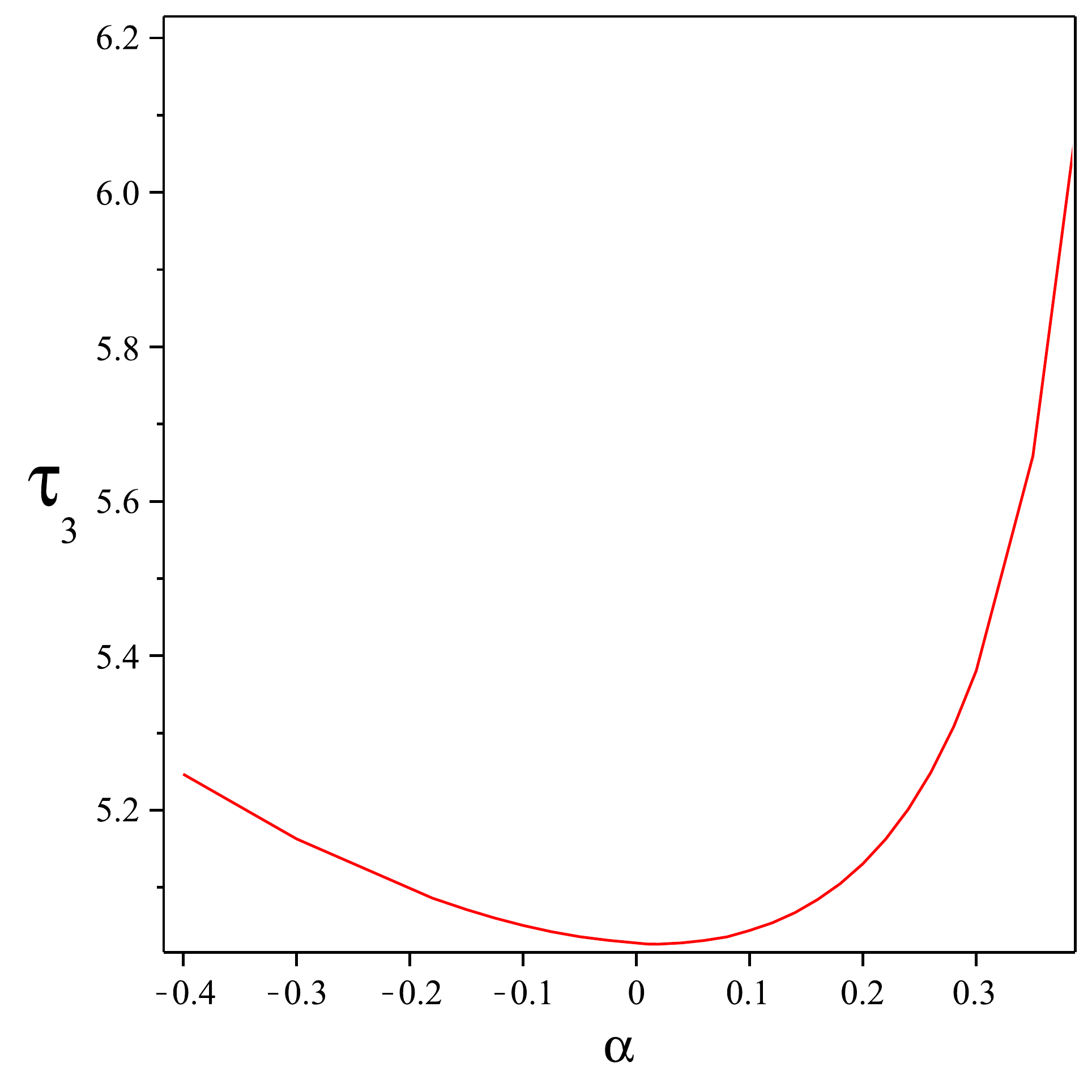}\quad\qquad\qquad}
{\includegraphics[width=.48\textwidth]{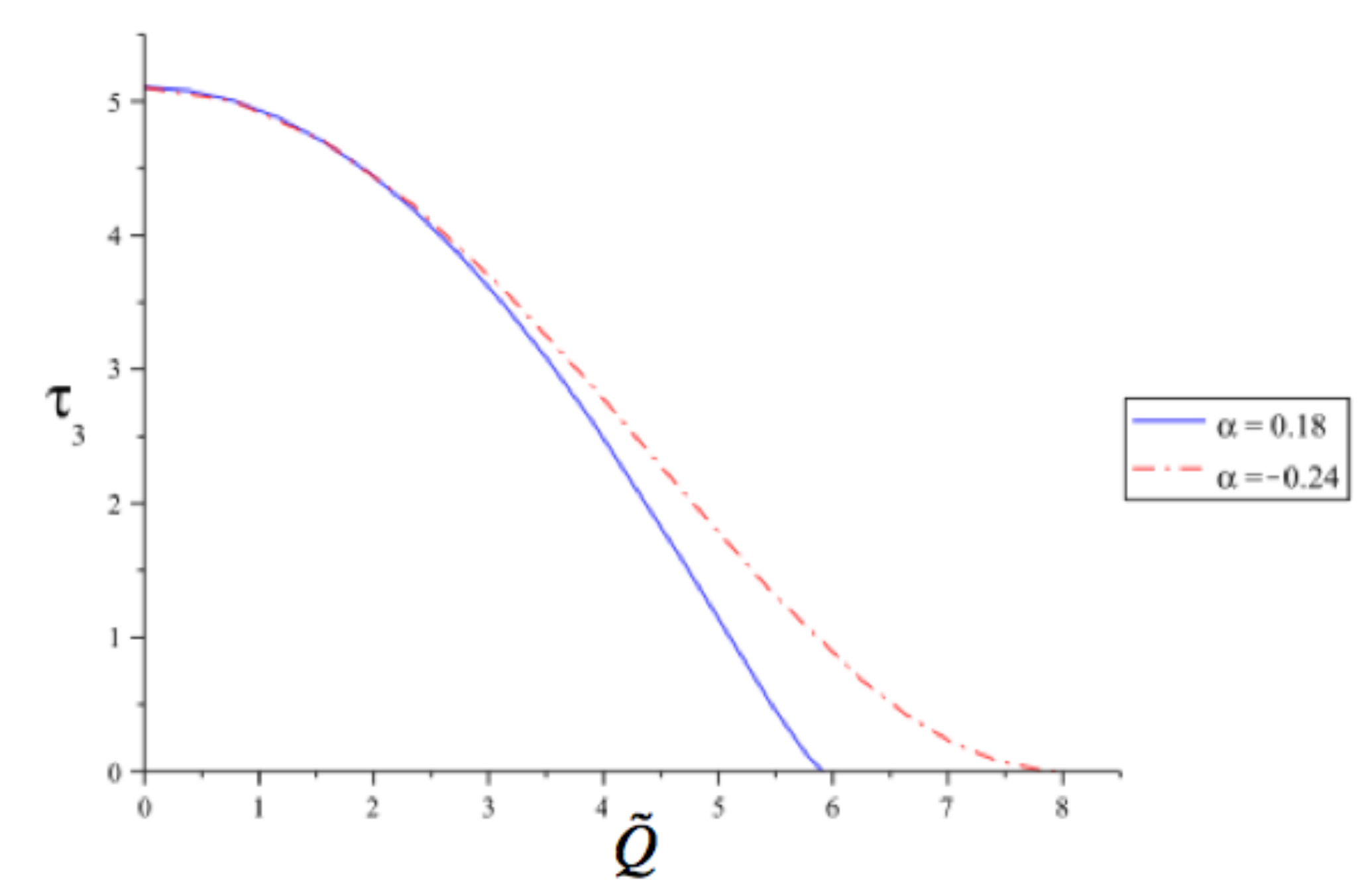}}
\caption{Left: $\tau_3$ vs $\alpha$ for 5-dimensional $z=3$
neutral black holes. Right: $\tau_3$ vs $\widetilde{Q}$ for
5-dimensional $z=3$ GB black holes with $\alpha=0.18$ (blue) and
$\alpha=-0.24$ (red) [$\alpha_m\approx0.02$]. } \label{tvsQz3}
\end{figure}

Fig. \ref{tvsQz2} illustrates this for $z=2$. The same situation occurs
for arbitrary values of $z>1$, as shown in fig. \ref{tvsQz3} for
$z=3$. We conclude that this is a general feature of planar
Gauss-Bonnet black holes of arbitrary size and $z>1$.

In general for a given $z$,  black holes with
$\alpha<\alpha_m$ show non-linear behaviour at low temperature. As
$\alpha$ decreases below $\alpha_m$, the maximum temperature,
$T(Q=0)$, increases as does $Q_{\mathrm{ext}}$. As a result the
non-linearity at low temperature becomes more significant as
$\alpha/\alpha_m$ becomes smaller. However as $\alpha$ increases
above $\alpha_m$, the maximum  black hole temperature increases
while $Q_{\mathrm{ext}}$ decreases.  In other words the temperature falls
faster, with the low temperature behaviour becoming increasingly linear as shown in fig. (\ref{tvsQz2}-right).

We close this section by noting that for $\alpha=\alpha_m$ no
isothermal pairs exist.

\subsection{Specific heat and Fermi liquid behaviour}

The specific heat at fixed volume is given by:
\begin{equation}
C=T\frac{(dS/dr_0)}{(dT/dr_0)}
\label{CT}
\end{equation}
where
\begin{equation}
S=\frac{1}{2}\frac{\pi^{n/2}r_0^{n-1}}{\Gamma(\frac{n}{2})}\label{antro}
\end{equation}
is the Bekenstein-Hawking entropy of the black brane.
\begin{figure}[tbp]
\centering
{\includegraphics[width=.8\textwidth]{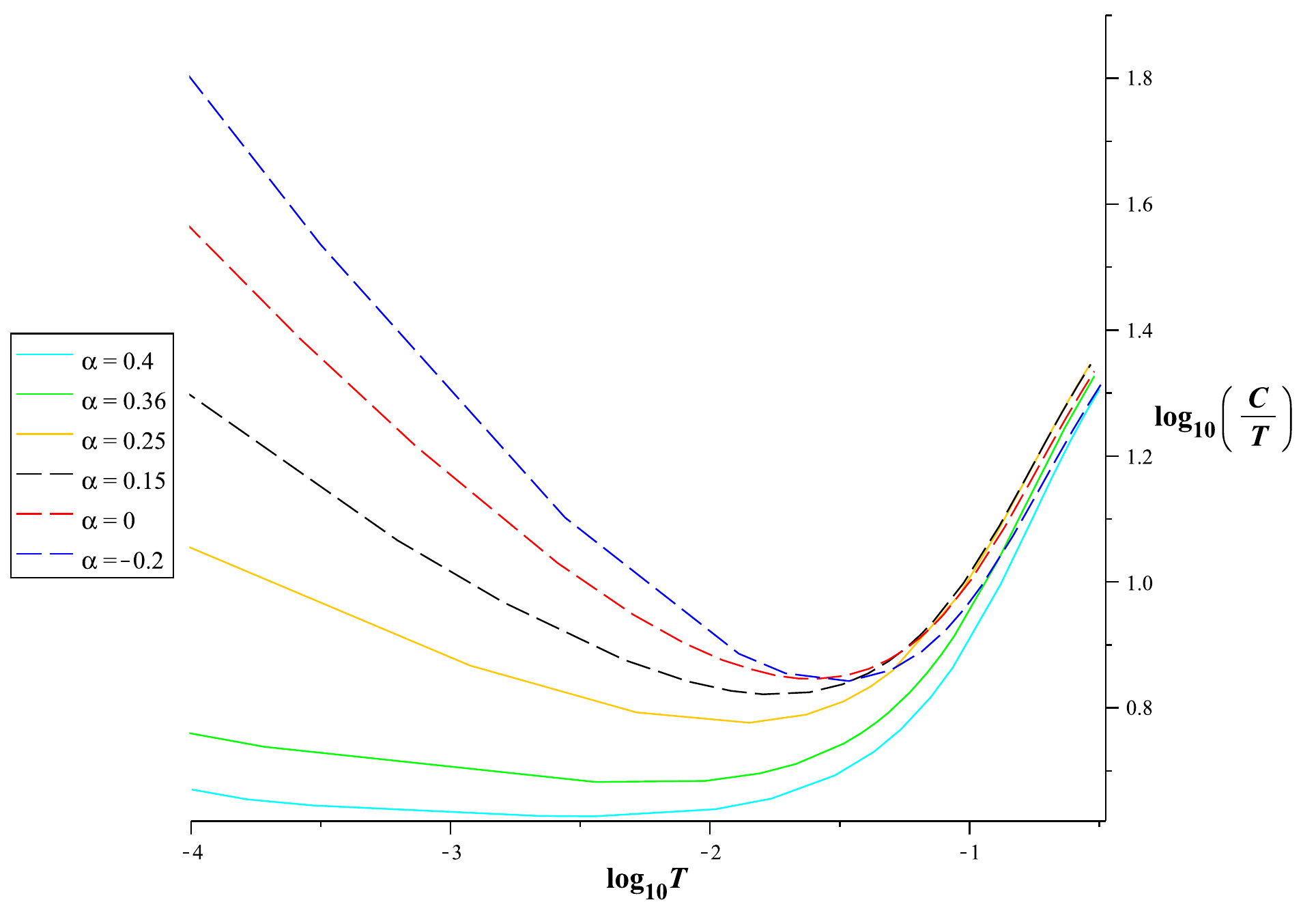}}\caption{$\log_{10}(C/T)$
vs $\log_{10}T$ in 5 dimensions  with fixed charge $Q=1$, 
$z=2$, and
$\alpha_m\approx0.22$.} \label{CTvsT}
\end{figure}
A characteristic behaviour of a Fermi liquid is that its specific heat
depends linearly on temperature.  A recent study of Fermi liquid holography compared
asymptotically AdS charged Einstein black holes($z=1$) to asymptotically
Lifshitz ones ($z>1$) \cite{Daniel2}. The former obeyed Fermi liquid behaviour at low temperature,
whereas the latter did not. Instead asymptotically
Lifshitz Einstein-Maxwell black holes ($z>1$) at low temperature
deviate from Fermi liquid behaviour. Indeed, for this class of
black holes $C/T$ diverges in the $T\rightarrow0$ limit.

To see how the GB correction modifies specific heat, we begin with
the $z=1$ case. Here we can easily obtain an analytic expression
for the specific heat
\begin{equation}
\frac{C}{T}=\frac{4(n-1)^2\pi^{\frac{n}{2}+1}r_0^{n-2}}%
{\Gamma(\frac{n}{2})\left[(2n-3)\widetilde{Q}^2+2n(n-1)(1-\frac{\alpha}{\ell^2})\right]}\label{CTz1}
\end{equation}
using eqs. (\ref{temp}), (\ref{tempz1}) and (\ref{antro}). From
eq. (\ref{CTz1}) it is straightforward to show that at high
temperature (i.e. $\widetilde{Q}\rightarrow0$) the Sommerfeld
ratio $C/T\propto T^{n-2}$. However at low temperature, where the
electric charge approaches the extremal value
$\widetilde{Q}_{\mathrm{ext}}(=Q_{\mathrm{ext}}/r_0^{n-1})$
introduced in (\ref{Qext}), the Sommerfeld ratio becomes constant:
\begin{equation}
\frac{C}{T}\rightarrow\frac{\pi^{\frac{n}{2}+1}r_{\mathrm{ext}}^{n-2}}%
{n\Gamma(\frac{n}{2})(1-\frac{\alpha}{\ell^2})}\qquad\mathrm{as}\qquad
T\rightarrow0
\end{equation}
with
\begin{equation}
r_{\mathrm{ext}}=\left[\frac{Q^2}{2n(n-1)(1-\frac{\alpha}{\ell^2})}\right]^{\frac{1}{2n-2}}
\end{equation}
The low-temperature constancy of the
Sommerfeld ratio is a key characteristic of Fermi liquids.

We have already observed that $T\propto r_0^z$ for arbitrary values
of $z$ in the
high temperature ($\widetilde{Q}\rightarrow0$) limit. Consequently, since
specific heat is proportional to $r_0^{n-1}$,  we find that
$C\propto T^{(n-1)/z}$ for all GB black holes.

The  low temperature limit  $C/T$ for $z>1$ must be obtained
numerically. We have numerically computed the specific heat
(\ref{CT}) for a range of values of $\alpha$. In fig. \ref{CTvsT}
we plot the Sommerfeld ratio $C/T$ as a function of temperature on
a log-log scale for $z=2$ for the isothermal pairs in fig.
\ref{tvsQz2}.  We see although both $\alpha<\alpha_m$ and
$\alpha>\alpha_m$ are isotherm pairs at high temperature, dramatic
growth appears in $C/T$ as $T\rightarrow0$ for $\alpha <\alpha_m$,
increasingly so as $\alpha$ becomes much smaller than $\alpha_m$.
Therefore the low temperature behaviour becomes increasingly non-linear
for GB black holes with $\alpha<\alpha_m$,  with  the Sommerfeld ratio
depending on temperature as a weak power law, consistent with the prediction of the
Castro Neto et al. model.  The
slope of the line in the log-log plot of $C_V/T$ at low temperature  from the bulk depends
 on the value of $\alpha$, whereas for the Castro Neto et al. model it is
characterized by $\lambda-1$. For example, for $\alpha=-0.2$ the
slope is approximately $-0.43$ (corresponding to $\lambda=0.57$) while it is approximately $-0.38$ for $\alpha=0$
(equivalent to $\lambda=0.62$).

The  $\alpha > \alpha_m$ isotherm counterparts exhibit rather
different behaviour at low temperature,  with the
$T$-dependence of the Sommerfeld ratio showing vanishingly small power law behaviour as
$\alpha$ increases (corresponding to $\lambda\rightarrow1$). For
$\alpha>>\alpha_m$, this becomes almost constant, restoring Fermi
liquid behaviour.

\section{Closing remarks}
\label{cm}

Adding a Gauss-Bonnet term to the Einstein-Hilbert action results
in interesting low temperature behaviour for asymptotically
Lifshitz charged black holes. By investigating the thermodynamic
properties of such black holes at low temperature, we found that
the Sommerfeld ratio $\gamma_0$ is temperature dependent,
 governed by a weak power law behaviour depending
on the value of $\alpha$. This kind of behaviour is a
characteristic property of some heavy fermion metals such as
f-electron compounds at low temperature. A CM model of these
non-Fermi liquids \cite{CNeto1, CNeto2} predicts weak power law
behaviour for other properties such as heat capacity and
susceptibility, i.e. $\gamma_0(T)\equiv C(T)/T\propto
\chi(T)\propto T^{-1+\lambda}$ with $\lambda<1$ .

 We propose that Lifshitz-GB charged black holes are gravitational duals to the Castro Neto et
al. model \cite{CNeto1, CNeto2}, with the Gauss-Bonnet coefficient
$\alpha$ holographically dual to the characteristic parameter
$\lambda$. To more firmly establish this duality will involve a
study of other thermodynamic quantities like resistivity or
susceptibility using the gravity theory. We leave this topic for
future work.

\section*{Acknowledgements}
This work was supported by the Natural Sciences and Engineering
Research Council of Canada.

\end{document}